\documentclass{kluwer}    
\usepackage[dvips]{graphicx}

\newdisplay{guess}{Conjecture}

\begin{document}                                                    
                               
\begin{article}
\begin{opening}         
\title{Modelling the Formation of Individual Galaxies: \\
A Morphology Problem for CDM?} 
\author{James~E. \surname{Taylor}\email{jet@astro.ox.ac.uk}}  
\institute{Astrophysics, University of Oxford\\ 
Denys Wilkinson Building, 1 Keble Road, Oxford OX1 3RH, United Kingdom}
\author{Arif \surname{Babul}\email{babul@uvic.ca}\thanks{CITA Senior Fellow}}  
\institute{Department of Physics and Astronomy, University of Victoria\\ 
Elliott Building, 3800 Finnerty Road, Victoria, BC, V8P 1A1, Canada}
\runningauthor{J.E. Taylor \& A. Babul}
\runningtitle{The Formation of Individual Galaxies}

\begin{abstract}
We use a semi-analytic model of halo formation to study 
the dynamical history of giant field galaxies like the Milky Way.
We find that in a concordance LCDM cosmology, most isolated disk galaxies
have remained undisturbed for 8--10 Gyr, such that the age of the Milky
Way's thin disk is unremarkable. Many systems also have older disk components 
which have been thickened by minor mergers, consistent with recent 
observations 
of nearby field galaxies. We do have a considerable problem, however, 
reproducing the morphological mix of nearby galaxies. In our fiducial
model, most systems have disk-to-bulge mass ratios of order 1, and look 
like S0s rather than spirals. This result depends mainly on merger statistics,
and is unchanged for most reasonable choices of our model parameters. 
We discuss two possible solutions to this morphology problem in LCDM.
\end{abstract}

\end{opening}           

\section{Introduction}  
\vspace{-1pc}
Recent numerical and semi-analytic models of galaxy formation
have made great progress in describing the average properties 
of populations of galaxies. Given the complexity of these models,
however, it is still not clear whether they are based on a
complete and definitive picture of galaxy formation. 
Here we report on preliminary results from a project to test
some basic components of galaxy formation models, 
using galaxy morphology.

We have used the semi-analytic model of Taylor (\shortcite{t01}, 
see also \inlinecite{tb01}) to produce individual histories of 
mass assembly and dynamical evolution for a representative set of
CDM galaxy halos. By combining this dynamical framework
with a simple prescription for galaxy formation, we can simulate 
the growth of individual galaxies at negligible computational 
expense. The systems considered in this work are a set of isolated field
galaxies like the Milky Way, in a concordance $\Omega = 0.3$, 
$\Lambda = 0.7$ LCDM cosmology.

\section{Disk Ages} 
\vspace{-1pc}
\noindent{\bf Thick Disks:} We assume that a 
collision with a large satellite (half the mass of the disk or more) 
will disrupt the disk forming at the centre of the main halo. A 
cumulative distribution of 
the resulting disk ages is shown in the left-hand panel of Fig.\ 1.

\noindent{\bf Thin Disks:} Minor collisions may
heat the thin disk, transforming it into a thick disk. 
In the right-hand panel of Fig.\ 1 
we show the distribution of thin disk age limits calculated 
in this way. We find that many systems have thin and thick 
disk components like those observed in nearby galaxies \cite{db02}.

\noindent{\bf Progressive Disk Heating:} We can also estimate
the progressive heating of stars as a function of age. In Fig.\ 2 
we compare the 
results of individual semi-analytic realisations with the observed 
age-dispersion relation 
for the Milky Way, from \inlinecite{qg01}.

\begin{figure}[h]
\tabcapfont
\centerline{
\begin{tabular}{c@{\hspace{6pc}}c}
\includegraphics[width=1.8in]{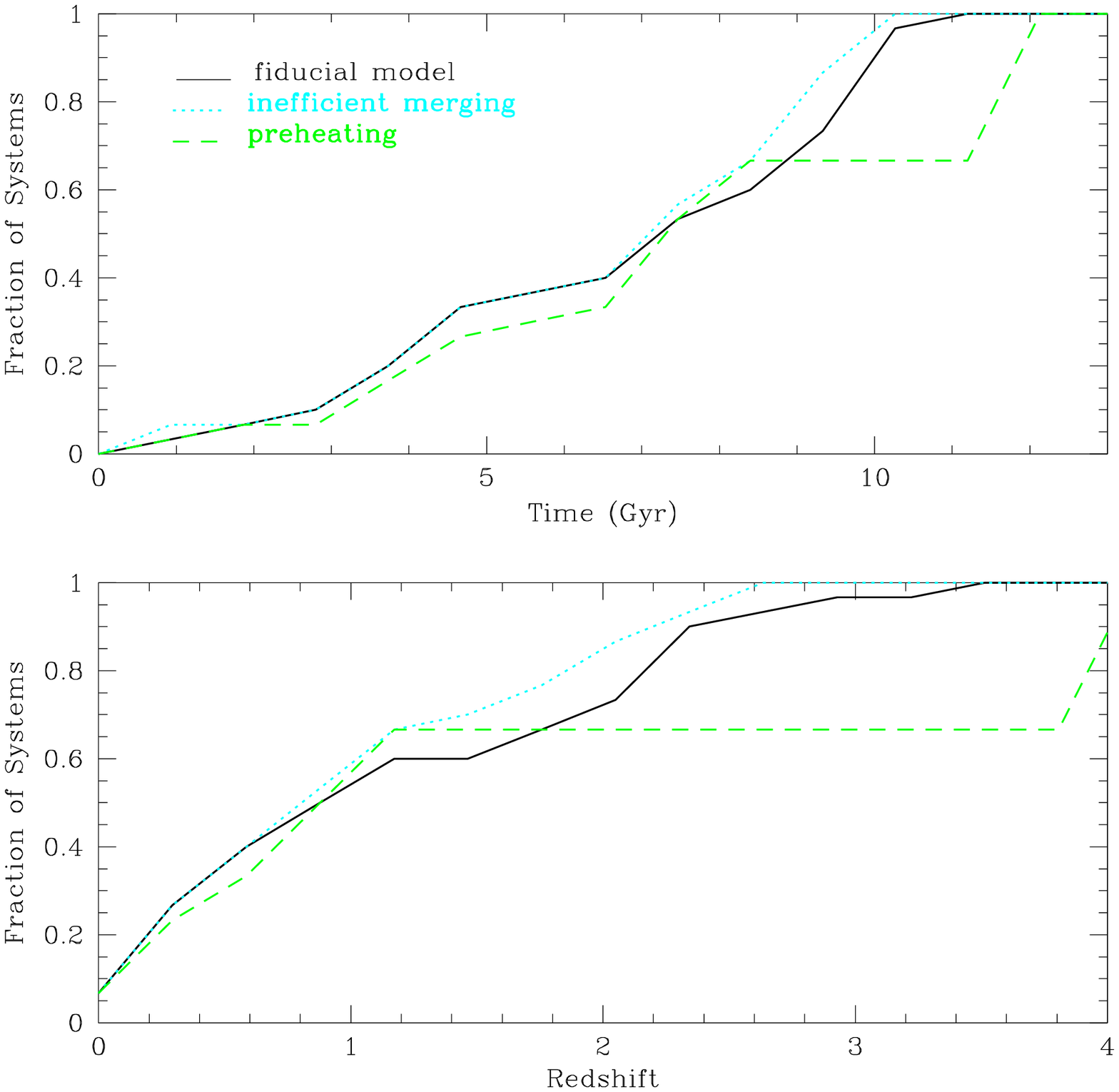} &
\includegraphics[width=1.8in]{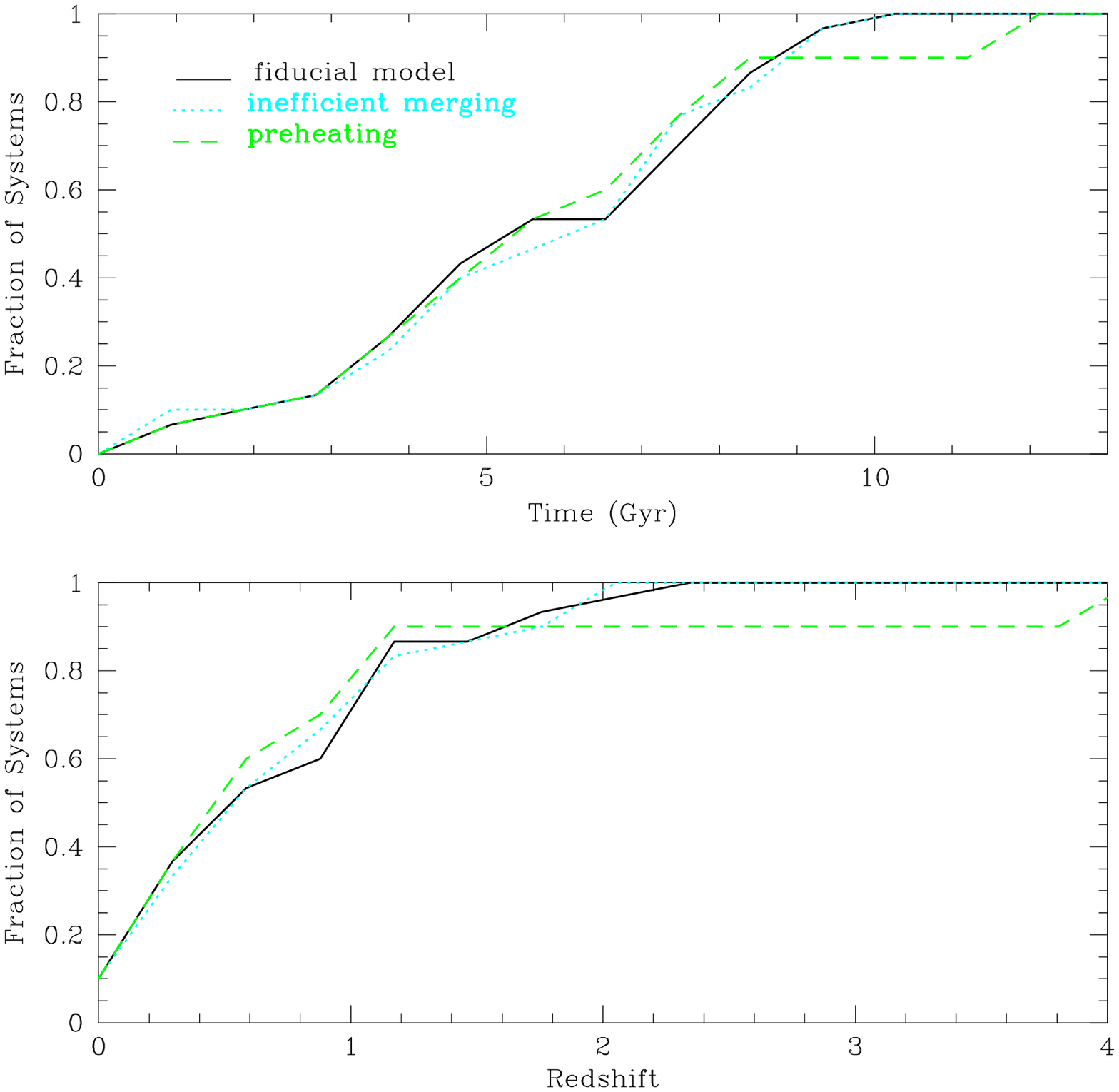}
\end{tabular}}
\caption{(Left) The cumulative age and redshift distributions of the last 
major merger, which marks the onset of the formation of the present thick 
disk. (Right) Same as left panel, but for minor mergers, which set the
age of the thin disk.}\label{fig:4-5}
\end{figure}

\vspace{-1pc}
\begin{figure}[b]
\centerline{\includegraphics[width=10pc]{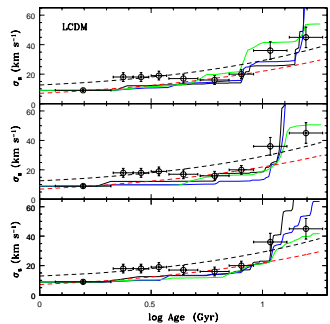}}
\caption{ The predicted increase in velocity dispersion 
from indirect heating (solid lines), compared with the observed age-dispersion 
relation (points, from Quillen \& Garnet (2001)), and smooth $t^{1/2}$ 
relations (dashed lines), for three different systems.}\label{fig:6}
\end{figure}

\vspace{-3pc}
\section{A Morphology Problem?}
\vspace{-1pc}
Given merger histories for our halos, we add a simple toy model
of galaxy formation to predict the growth of the main components of 
the central galaxy. 
Hot gas in the halo cools on the infall timescale and is added to the disk,
disk material is partially transferred to the spheroid in disruption events,
and disrupted satellites are also added to the spheroid (Fig.\ 3). 
This model is not designed
to provide a detailed account of star formation within each system, but rather
to place gross constraints on when the components of field galaxies could 
have formed.

Examining the histories of these systems, we notice a
trend: massive disks form early and are often disrupted and converted into
bulges, producing bulge-dominated morphologies at the present day. While the
conversion from mass ratios to galaxy morphologies is problematic, we can
indicate the effect schematically by plotting the two components in each
system, scaled by their mass (Fig.\ 4). 

\begin{figure}[h]
\tabcapfont
\centerline{
\begin{tabular}{c@{\hspace{3pc}}c}
\includegraphics[width=1.6in]{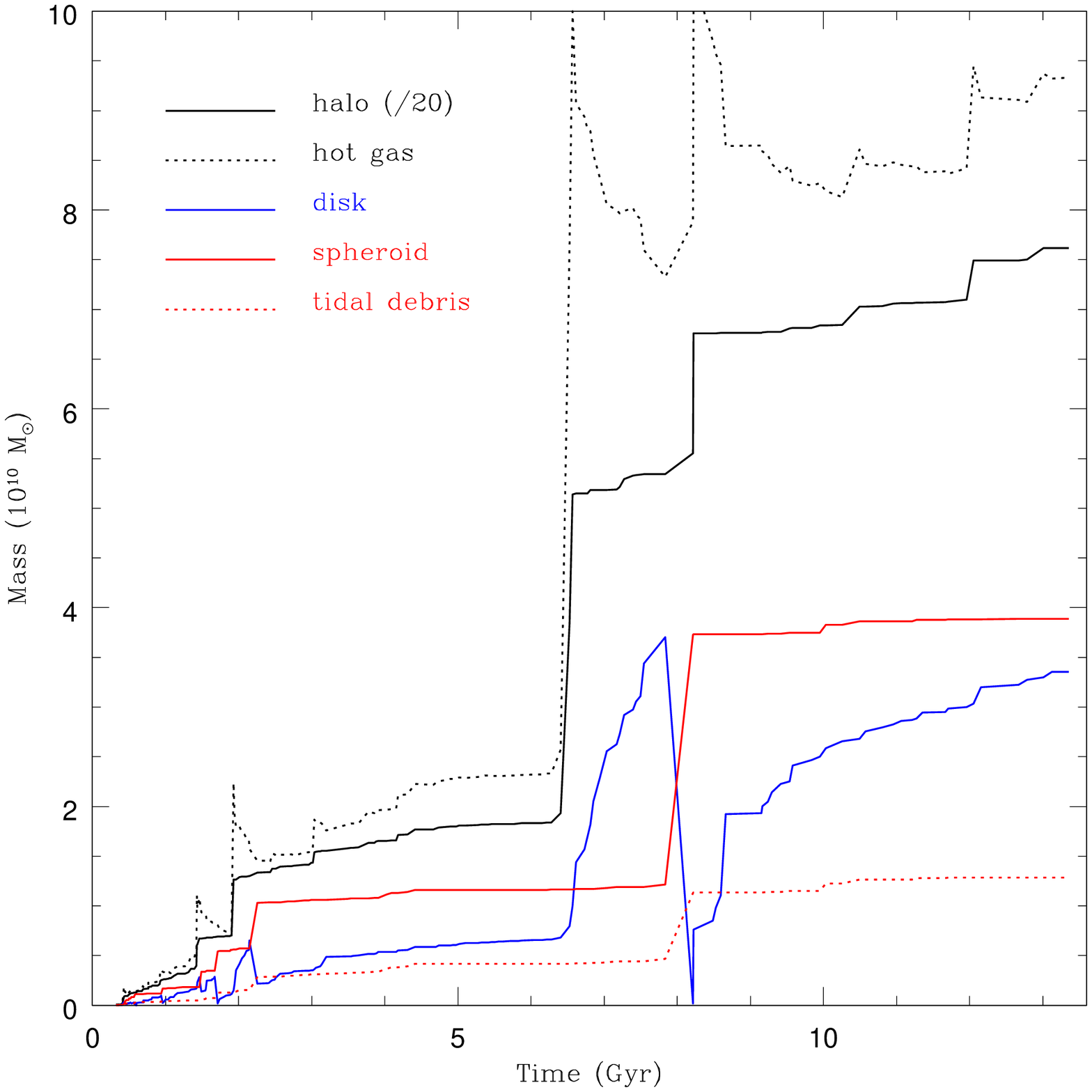} &
\includegraphics[width=1.6in]{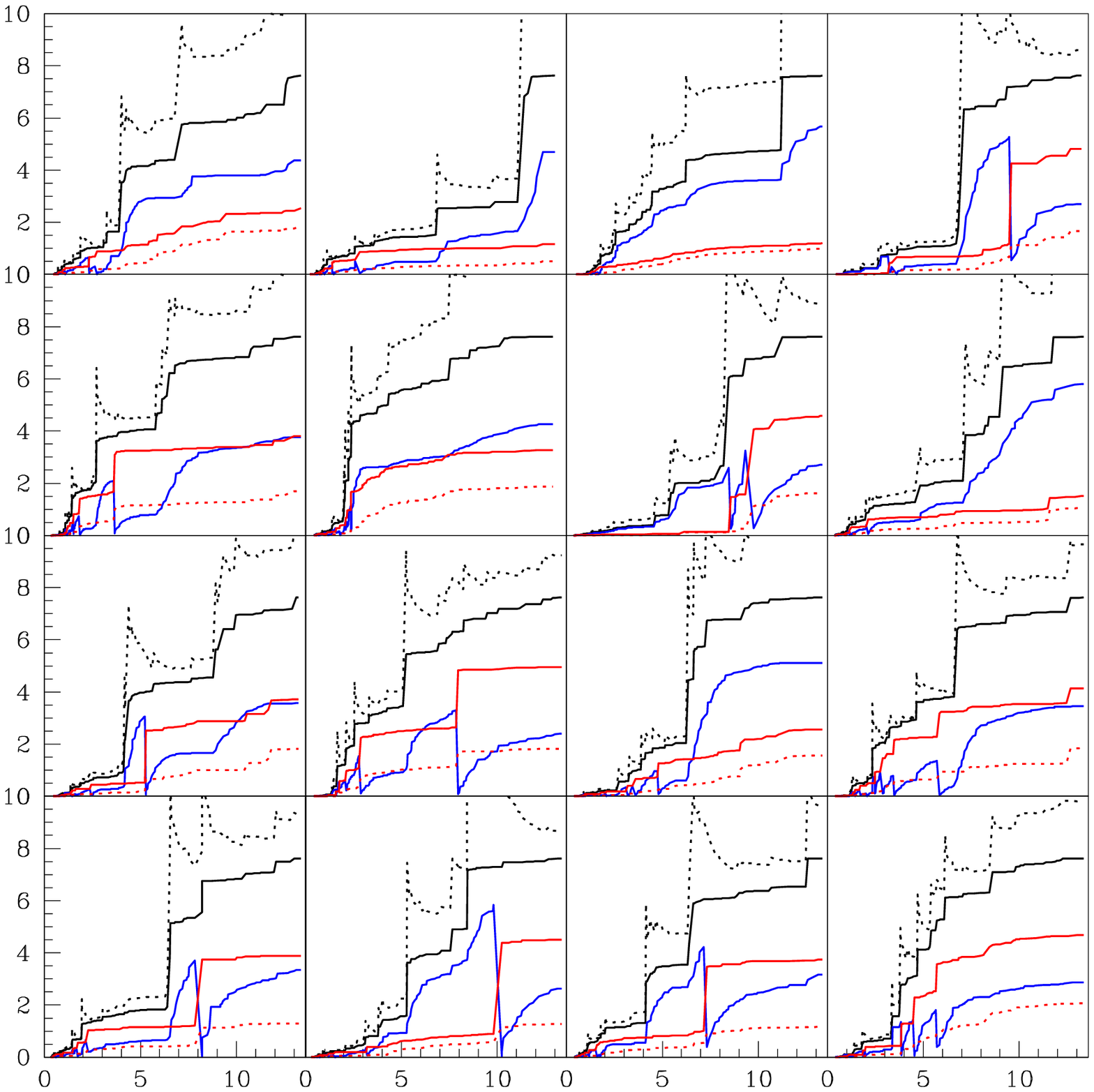}
\end{tabular}}
\caption{(Left) The growth of the main components in a single 
system, showing the effects of cooling and disk disruption.
 (Right) The same for a large set of systems.}\label{fig:7-8}
\end{figure}

\vspace{-1pc}
\begin{figure}[b]
\centerline{\includegraphics[width=11pc]{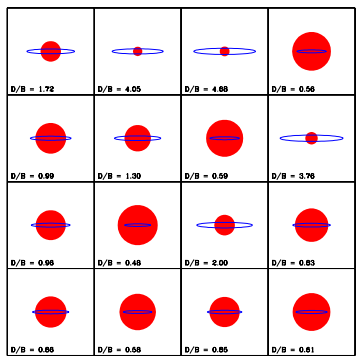}}
\caption{Schematic morphologies for some of our 
fiducial systems.
The spheroids and disks are scaled by the mass of the components
in each system, and the disk-to-bulge mass ratio is indicated.}\label{fig:9}
\end{figure}

\vspace{-2pc}
\section{Possible Solutions}
\vspace{-1pc}
\noindent{\bf Inefficient Starbursts?} In our fiducial 
model, we assumed that 50\% of the disk mass is transferred to the spheroid in 
the average disruption event, as old disk stars are scattered out of the plane 
and new stars form in an extended starburst. If high-redshift disks are 
gaseous and
starbursts are inefficient, this conversion rate may be lower and 
the average disk-to-bulge ratio can increase, as shown in Fig.\ 5.

\noindent{\bf Gas Preheating?} Another possibility
is that disk formation is delayed, possibly by some extra source of heat
or entropy in the halo gas at early times. If we prevent all cooling prior to
z = 1.5, we get a better mix of morphologies (Fig.\ 5, right panel), 
but this solution may
conflict with observations of large disks at high redshift.

\begin{figure}[h]
\tabcapfont
\centerline{
\begin{tabular}{c@{\hspace{3pc}}c}
\includegraphics[width=1.5in]{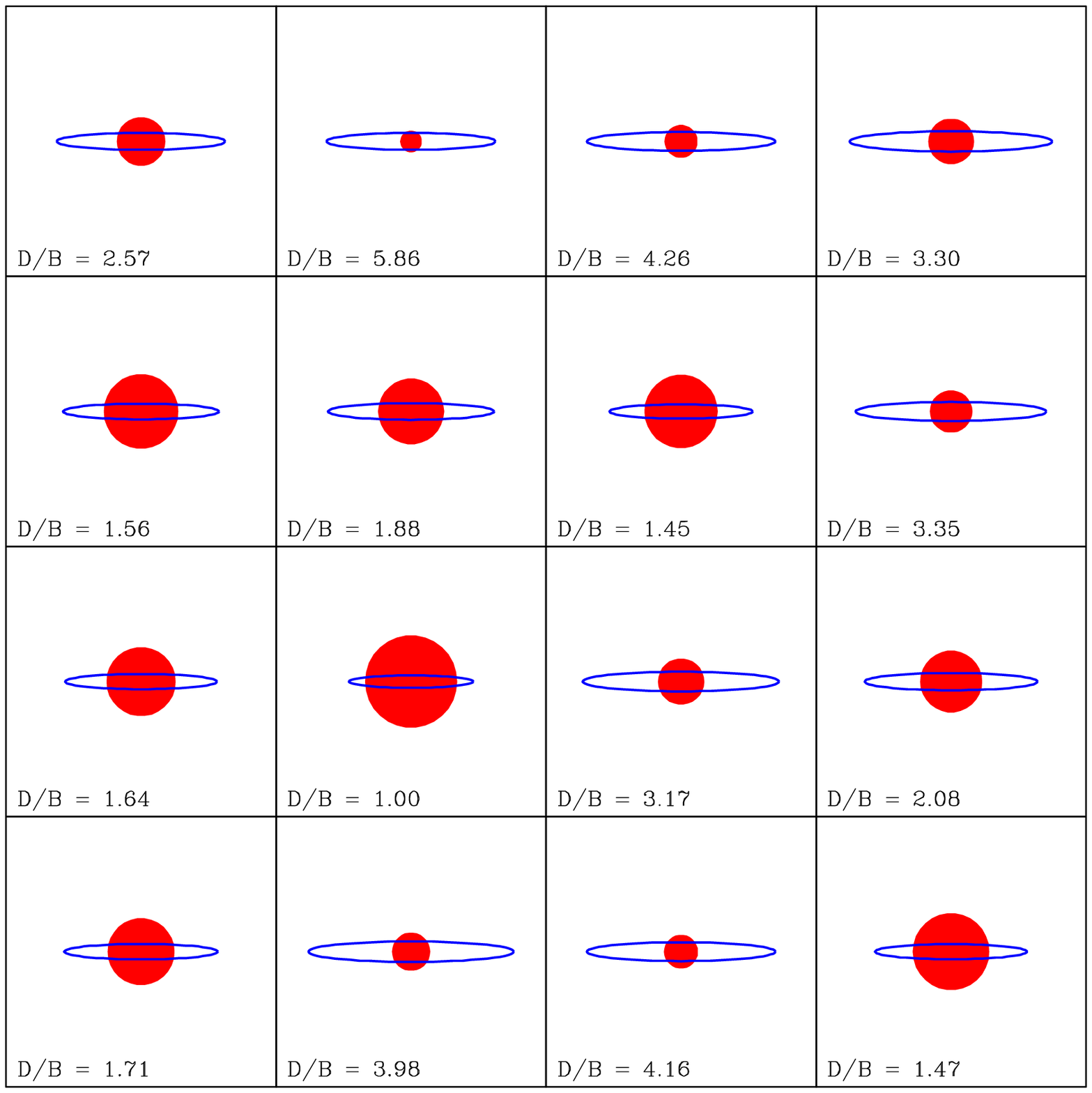} &
\includegraphics[width=1.5in]{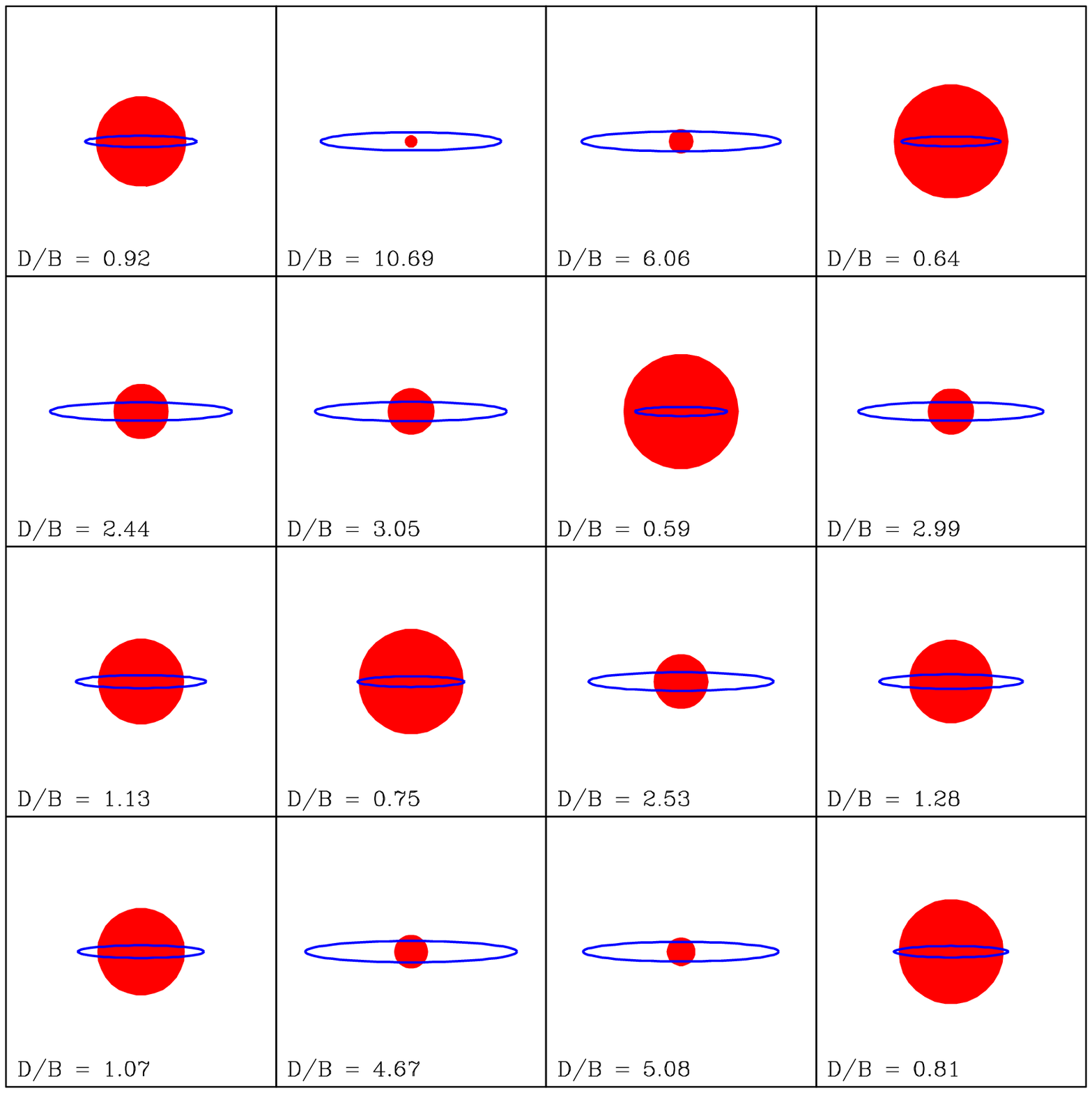}
\end{tabular}}
\caption{(Left) Schematic morphologies for systems assuming disruptions
require a 1:1 collision with a satellite, and transfer only 
25\% of the disk mass 
to the spheroid. (Right) Schematic morphologies for systems where cooling 
is delayed until z = 1.5.}\label{fig:10-11}
\end{figure}

\vspace{-2pc}
\section{Conclusion} 
\vspace{-1pc}
Examining the dynamical history of galaxy halos in a LCDM cosmology,
we find that
while the centres of most halos have remained undisturbed for many Gyr, 
the existence of old, thin and {\it massive} components 
in the disks of galaxies like the Milky Way is still a problem for LCDM, 
as such systems form early and lose most of their mass through successive 
disruptive encounters at early times. Thus, unless disks are 
exceptionally robust, there must be some mechanism which delays cooling 
and star formation in massive halos at early times. The `preheating' 
mechanism which appears to produce an entropy floor in cluster gas \cite{be02}
may be one possible candidate.

\end{article}
\end{document}